    \documentclass[number]{elsarticle}

    \newcounter{bla}
    \newenvironment{refnummer}{%
    \list{[\arabic{bla}]}%
    {\usecounter{bla}%
     \setlength{\itemindent}{0pt}%
     \setlength{\topsep}{0pt}%
     \setlength{\itemsep}{0pt}%
     \setlength{\labelsep}{2pt}%
     \setlength{\listparindent}{0pt}%
     \settowidth{\labelwidth}{[9]}%
     \setlength{\leftmargin}{\labelwidth}%
     \addtolength{\leftmargin}{\labelsep}%
     \setlength{\rightmargin}{0pt}}}
     {\endlist}

    \usepackage{natbib}
    \usepackage{hyperref}
    \usepackage{url}
    \usepackage[pdftex]{graphicx,color}
    \usepackage{epsfig}
    \usepackage{epstopdf}
    \usepackage{color}
    \usepackage{ifthen}
    \newcounter{showchanges}
    \newcounter{paperVersion}
    \newcommand{\onlyShortVersion}[2][]{\ifthenelse{\value{paperVersion}=0}{#2}{#1}}
    \newcommand{\onlyLongVersion}[2][]{\ifthenelse{\value{paperVersion}>0}{#2}{#1}}
    \def\addWide#1{}

    \newcommand{\ourtitle}{HONEI: A collection of libraries for numerical computations targeting multiple processor architectures}
\newcommand{\ourkeywords}{High performance computing; FEM for PDE; Shallow Water Equations; mixed precision methods; CUDA; Cell BE}
\newcommand{\ourpacscodes}{02.70.-c (Computational techniques; simulations), 07.05.Bx (Computer systems: hardware, operating systems, computer languages, and utilities), 89.20.Ff (Computer Science and Technology), 47.11.-j (Computational methods in fluid dynamics)}

\begin{document}
    \begin{frontmatter}

    \title{\ourtitle}

    \author[physics]{Danny van Dyk}
    \author[math]{Markus Geveler}
    \author[cs]{Sven Mallach}
    \author[math]{Dirk Ribbrock}
    \author[math]{Dominik G\"oddeke\corref{cor}\fnref{dfgdom}}
    \author[cs]{Carsten Gutwenger}

    \cortext[cor]{Corresponding author}
    \fntext[dfgdom]{Supported by DFG, project TU102/22-1, TU102/22-2}

    \address[physics]{Institut f\"ur Physik, TU Dortmund}
    \address[math]{Angewandte Mathematik, TU Dortmund}
    \address[cs]{Informatik, TU Dortmund}

    \begin{abstract}
    We present HONEI, an open-source collection of libraries offering a hardware oriented approach to numerical calculations. HONEI abstracts the hardware, and applications written on top of HONEI can be executed on a wide range of computer architectures such as CPUs, GPUs and the Cell processor. We demonstrate the flexibility and performance of our approach with two test applications, a Finite Element multigrid solver for the Poisson problem and a robust and fast simulation of shallow water waves.
    By linking against HONEI's libraries, we achieve a twofold speedup over straight forward C++ code using HONEI's SSE backend, and additional 3--4 and 4--16 times faster execution on the Cell and a GPU. A second important aspect of our approach is that the full performance capabilities of the hardware under consideration can be exploited by adding optimised application-specific operations to the HONEI libraries. HONEI provides all necessary infrastructure for development and evaluation of such kernels, significantly simplifying their development.
    \begin{flushleft}
    PACS: \ourpacscodes
    \end{flushleft}

    \begin{keyword}
\ourkeywords
    \end{keyword}

    \end{abstract}

    \end{frontmatter}
\ifnum 0=1
\newpage

    {\bf PROGRAM SUMMARY}

    \begin{small}
    \noindent
    {\em Manuscript Title: \ourtitle}                             \\
    {\em Authors: D.van Dyk, M.Geveler, S.Mallach, D.Ribbrock, D.G\"oddeke and C.Gutwenger}                                                \\
    {\em Program Title: HONEI}                                          \\
    {\em Journal Reference:}                                      \\
    {\em Catalogue identifier:}                                   \\
    {\em Licensing provisions: GPLv2}                                   \\
    {\em Programming language: C++}                                   \\
    {\em Computer: x86, x86\_64, NVIDIA CUDA GPUs, Cell blades and PlayStation 3}                                               \\
    {\em Operating system: Linux}                                       \\
    {\em RAM:} at least 500 MB free                                    \\
    {\em Number of processors used: 1 x86 core (plus one GPU, or one Cell Processor)}                              \\
    {\em Supplementary material:}                                 \\
    {\em Keywords:} \ourkeywords  \\
    {\em PACS:} \ourpacscodes                   \\
    {\em Classification:}  4.8 (Linear Equations and Matrices), 4.3 (Differential Equations), 6.1 (Hardware)                                       \\
    {\em External routines/libraries: } SSE: none; [1] for GPU, [2] for Cell backend  \\
    {\em Nature of problem:}  \\
Computational science in general and numerical simulation in particular have reached a turning point. The revolution developers are facing is not primarily driven by a change in (problem-specific) methodology, but rather by the fundamental paradigm shift of the underlying hardware towards heterogeneity and parallelism. This is particularly relevant for data-intensive problems stemming from discretisations with local support, such as finite differences, volumes and elements.
\\
    {\em Solution method:}\\
To address these issues, we present a hardware aware collection of libraries combining the advantages of modern software techniques and hardware oriented programming. Applications built on top of these libraries can be configured trivially to execute on CPUs, GPUs or the Cell processor. 
In order to evaluate the performance and accuracy of our approach, we provide two domain specific applications; a multigrid solver for the poisson problem and a fully explicit solver for 2D shallow water equations.
\\
    {\em Restrictions:}\\
    HONEI is actively being developed, and its feature list is continuously expanded. Not all combinations of operations and architectures might be supported in earlier versions of the code. Obtaining snapshots from \url{http://www.honei.org} is recommended.
    \\
{\em Unusual features:}\\
The considered applications as well as all library operations can be run on NVIDIA GPUs and the Cell BE.
   \\
{\em Running time:}\\
Depending on the application, and the input sizes. The Poisson solver executes in few seconds, while the SWE solver requires up to 5 minutes for large spatial discretisations or small timesteps.
   \\
{\em References:}
\begin{refnummer}
\item \url{http://www.nvidia.com/cuda}     
\item \url{http://www.ibm.com/developerworks/power/cell}            
\end{refnummer}
\end{small}

\newpage
\hspace{1pc}
{\bf LONG WRITE-UP}

\fi

\section{Introduction}\label{sec:introduction}

Computational science in general and numerical simulation in particular have reached 
a turning point. The revolution developers are facing is not primarily driven by a 
change in (problem-specific) methodology, but rather by the fundamental paradigm shift 
of the underlying hardware towards heterogeneity and parallelism.

\subsection{Hardware}\label{sec:intro:hw}

The general expectation that performance of serial codes improves automatically is 
no longer true. Power and heat considerations have put an end to aggressively 
optimising circuits and deeper processor pipelines (\emph{power wall}). 
\emph{Frequency scaling} is thus limited to \lq natural frequency scaling\rq\ by process 
shrinking. Ever wider superscalar architectures with more aggressive out-of-order execution 
have reached their limits, the hardware is no longer able to extract enough instruction-level 
parallelism to hide latencies (\emph{ILP wall}).

The primary solution provided by processor manufacturers is the migration towards 
on-chip parallelism in the form of chip multiprocessors (CMP); the (approximately) 
doubled amount of transistors per chip generation is invested into a doubling of 
processing cores. Over the last few years, both Intel and AMD have established 
multicore processors in the mass market.

At the same time, memory performance continues to improve at a significantly slower 
rate than compute performance. In particular due to pin limits, this \emph{memory wall} problem is 
even worsened by multicore architectures. Top-end quadcore processors can achieve a 
theoretical peak floating point performance of 12\,GFLOP/s per core in double precision (24\,GFLOP/s in single precision)
whereas the bandwidth to off-chip memory has barely increased over the last 
hardware generations, now reaching approximately 12\,GB/s shared between all cores on a die: Bandwidth 
asymptotically scales with the number of sockets and not with the number of cores. In the scope of this paper, the \emph{arithmetic intensity} (defined as the ratio of floating point operations on each data item) determines the attainable performance, ratios of 1:1 are not uncommon. The traditional approach to add ever larger on-chip 
cache memory hierarchies with sophisticated prefetching and replacement policies only 
alleviates the memory wall problem if data reuse is possible.

In the context of these three fundamental problems, alternative processor designs such 
as graphics processors (GPUs) or the Cell BE processor, which has been primarily 
developed for the gaming console PlayStation 3, are of particular interest. For instance, 
NVIDIA's flagship GPU at the time of writing, the GeForce GTX~285, features 30 
multiprocessors (each comprising 8 single precision \emph{streaming processors} (SP), 
one double precision SP, a shared lock-step instruction unit and support for transcedentals) 
delivering more than 1\,TFLOP/s theoretical peak performance in single precision, and more 
importantly, a sustained, achievable memory bandwidth of 160\,GB/s. This outstanding level 
of performance is achieved by keeping tens of thousands of threads \lq in flight\rq\ 
simultaneously to effectively hide latencies; context switches between threads are handled by the hardware at no additional cost. On the other hand, the Cell processor is an 
example of on-chip heterogeneity, comprising a standard Power 5 CPU (PPE) and eight on-chip 
floating point co-processors called \emph{Synergistic Processing Elements} (SPEs). The SPEs 
and the PPE are connected via a fast, 200\,GB/s on-chip ring bus, the theoretical peak performance 
is 230\,GFLOP/s in single precision, and off-chip memory is accessible at 25.6\,GB/s. With 
achievable performance levels significantly exceeding that of commodity designs, and a 
dramatic improvement of toolchain support in the past two years, such processors migrate  
from specialised solutions to the general-purpose computing domain and are 
widely considered as forerunners of future manycore designs.

\subsection{Software}\label{sec:intro:sw}

This parallelisation and heterogenisation of resources imposes a major challenge to the 
application programmer: Practitioners in computational science (numerics, natural sciences and engineering) often concentrate 
primarily on advancements in domain-specific methodology and hesitate to deal with the 
complications resulting from the paradigm shift in the underlying hardware. An important 
observation, that can be made in many existing software solutions, is that many algorithms 
are formulated---on the application level---to concentrate most of the actual work 
in few kernels, separated from the generic control flow. Examples include both low-level 
kernels such as matrix-vector products as well as high-level kernels such as black-box 
iterative solvers for sparse linear systems. In these cases, well-defined interfaces between 
the kernels enable individual tuning and, ideally, specialisation for different hardware 
resources. The idea to encapsulate the resulting highly tuned kernels into libraries is as 
old as scientific computing. 

For applications involving sparse rather than dense matrices and correspondingly a low 
arithmetic intensity, little standardisation exists and therefore, applications cannot 
typically be \lq plugged together\rq\ from calls to standard libraries. In particular, 
little standardisation exists in terms of data structures. 

Ultimately, compiler support to exploit on-chip parallelism efficiently is required. 
Experience with preliminary auto-parallelisation and auto-vectorisation (SIMD, SSE) 
features in modern compilers or OpenMP-like parallelism guided by hints given by the 
programmer, however, has taught us that compilers are still far from optimising code 
automatically, in particular codes limited by the memory wall problem (NUMA etc.). Compiler treatment 
of heterogeneous resources is still 
in its infancies, even though compilers (and associated languages) such as 
Sequoia~\citep{Fatahalian:2006:SPT}, tailored for memory hierarchies, promise better performance.

\subsection{Paper contribution}\label{sec:contribution}

Vendor-supplied BLAS, LAPACK and FFT libraries are available for multicore and emerging manycore architectures in the scope of this paper, for instance the highly tuned implementations of MKL (Intel CPUs), CUBLAS and CUFFT (NVIDIA GPUs), ACML-GPU (AMD multicore CPUs, AMD/ATI GPUs) and IBM's CBE BLAS, LAPACK and FFT libraries. However, all vendor-provided implementations are closed source, and currently do not yet support the full functional range of their (single-threaded) CPU counterparts.

We are convinced that in many practical cases, relying on library-based acceleration for common operations is insufficient, simply because little standardisation exists beyond the above mentioned approaches. Rather, it is unavoidable to implement at least a few specialised kernels to achieve application-specific functionality and a reasonably high performance simultaneously. In order to port such application-specific kernels to novel and heterogeneous architectures, application programmers need to learn both a new language to implement the actual kernels, and an associated runtime environment. The latter is responsible for scheduling computations, managing data transfers between the coordinating host CPU and the hardware accelerators; and even the physical data storage, e.g., by meeting appropriate alignment criteria and converting data structures on the fly.

In this paper we present HONEI (\emph{hardware-oriented numerics efficiently implemented}), a collection of architecture-aware libraries designed to alleviate the conflicting goals outlined above. HONEI is free software, and the existing libraries and infrastructural features are already available to the open source community. Our primary design goal is to abstract from architecture-related implementational details as much as possible. We addresses two major use case scenarios simultaneously:

\begin{itemize}
\item HONEI enables application-specific kernel development by providing all necessary architecture-dependent infrastructure in such a way that hardware details are dealt with automatically by a common runtime environment. As a consequence, the user can concentrate on the actual kernel implementation and is alleviated from writing e.g. data transfer operations or mailbox-code to synchronise low-level communication on the Cell processor.

\item HONEI automatically accelerates applications that are built entirely on top of its linear algebra operations. Its supported features, in particular with respect to architecture-aware optimisations, are continuously being augmented. Users only have to change an architecture tag in function calls (see Section~\ref{sec:overview_software}) to benefit from hardware acceleration.
\end{itemize}

\subsection{Related work and comparison with other libraries}\label{sec:relatedwork}

Our nomenclature of \emph{hardware-oriented numerics} is based on the work by Keyes 
and Colella et al., who survey trends towards terascale computing for a wide range 
of bandwidth-limited applications and conclude that only a combination 
of techniques from computer architecture, software engineering, numerical modeling and 
numerical analysis will enable a satisfactory and future-proof scale-out on the 
application level~\citep{Keyes:2002:TIM, Colella:2003:ASB}.

Many publications discuss detailed (implementational) aspects of the memory wall problem 
for multigrid solvers and sparse matrices in general: Douglas and Thorne, and Douglas, 
R\"ude et al.\ present cache-oriented multigrid solver components~\citep{Douglas:2002:ANO, 
Douglas:2000:FAA, Douglas:2000:COF}. Dongarra's group focusses on dense linear algebra: They examined mixed precision methods~\citep{Langou:2006:ETP}; and with the PLASMA project, they started research into optimised dense data structures for multicore architectures~\citep{Buttari:2006:TIO, PLASMA-web-page}.  Williams 
et al.\ and Owens et al.\ provide a detailed overview and many examples of scientific 
computations on the Cell processor and on GPUs~\citep{Williams:2006:TPO, Owens:2007:ASO, 
Owens:2008:GC}.

Publicly available academic software packages simultaneously aiming at numerical and hardware efficiency include PETSc, Trilinos,
DUNE and MTL4~\citep{petsc-web-page, Heroux:2005:AOO, Blatt:2008:OTG, Gottschling:2007:RTM}. A performance comparison 
with these packages is beyond the scope of this paper, and clearly not intended. HONEI emphasises hardware-orientation and provides both efficient implementations as well as infrastructure, not limited to
general-purpose CPUs. To the best of our knowledge, it is the only available set of libraries 
tuned for common CPUs, the Cell processor and GPUs simultaneously. In order to achieve this, HONEI is not build on top of
established other libraries, but has been designed from scratch. Its extensible structure dividing data and (architecture-optimised) computation
is continuously augmented. We are aware that other packages are (currently) more evolved, with a richer set of high level kernels and a much wider application basis for commodity based CPU (distributed memory) settings. 

The recent standardisation of OpenCL provides a common interface to heterogeneous multicore hardware. However, OpenCL is very low-level, and high-level approaches continue to be important for accessibility and usability reasons.

\subsection{Paper organisation}\label{sec:intro:paperorga}

Section~\ref{sec:background} provides the background of the two model applications used 
in this paper. In the subsequent three sections, HONEI's paradigms and structure 
are described in detail. Section~\ref{sec:results} finally presents numerical and benchmark 
results for the components and applications.

We deliberately omit code samples in this paper to improve readability and keep the presentation compact. Instead, 
the accompanying tarball as well as our homepage \url{http://www.honei.org} offer a detailed tutorial, demonstrating HONEI use cases and providing a more hands-on, code-oriented introduction to HONEI features and concepts. 


\section{Theoretical background}\label{sec:background}

This section describes the theoretical background of two example applications built on top of HONEI, that are used throughout this paper: The Poisson problem is a fundamental model problem as it appears as a subproblem in many different areas, for instance in electrostatics, solid mechanics and fluid dynamics. Solving the Shallow Water Equations (SWE) is of huge interest for many practical flow problems such as Tsunami simulations. Its implementation uses a wide variety of non-trivial operations and containers, justifying its choice as a demo application for the HONEI framework.

\subsection{The Poisson problem}\label{sec:poisson}

Our first test problem is the well-known Poisson equation on a 2D unitsquare domain $\Omega$:
\begin{equation}
-\Delta \mathbf{u} = \mathbf{f} 
\end{equation}
We apply mixed Dirichlet and Neumann boundary conditions, discretise the domain with bilinear conforming Finite Elements from the $Q_1$ space and enumerate the degrees of freedom in a generalised tensor product fashion. The resulting stiffness matrix has thus an a-priori known band structure, which can be exploited in linear algebra components via cache blocking~\citep{Turek:2004:HON}. A grid refinement level of $L$ yields $(2^L+1)^2$ unknowns, and we employ a (geometric) multigrid scheme to solve the associated sparse linear system. Even though it suffices to employ a matrix stencil for this simple grid, we still assemble the matrix fully, as this is the relevant case in practice due to grid and operator anisotropies.

\subsection{The 2D Shallow Water equations}\label{sec:swe}

Our second application is a fully explicit 2D shallow water simulation.  \onlyLongVersion{A basic feature of flows in rivers, reservoirs, harbours or coastal regions is, that they are characterised by horizontal motion (which means that the wavelength is usually much greater than its height). Due to this, the vertical velocity can be neglected in favour of a depth-averaged quantity, which leads to a simplification of the general 3D Navier Stokes equations of fluid flow.} The governing equations for the application---the 2D Shallow Water equations (SWE)---can be written as
\begin{equation}
 \label{eq:releg} \mathbf{U}_t + F(\mathbf{U})_x + G(\mathbf{U})_y = S(\mathbf{U}), \ \ (x,y) \in \Omega, \ \  t \ge 0, \\ \  \mathbf{U}= \left( h\ \ hu_1\ \ hu_2 \right)^T
\end{equation}

where $h$ is the water depth and $u_i$ are the velocities in $x$- and $y$-direction. $F$ and $G$ denote the flow in $x$- and $y$- direction, respectively.

We use the relaxation scheme proposed by Delis and Katsaounis with a second order accurate Finite Difference discretisation in space and a Runge Kutta time stepping mechanism~\cite{Delis:2005:DK2}. The solver is capable of computing numerically critical scenarios with low or even zero-valued initial water depth (dry states) and takes properties of the bed topography into account by applying a source term $S$. Space constraints prohibit a detailed description of the complex scheme, and we refer the reader to the original publication for details. Instead, we note that as the solver is explicit, no linear systems have to be solved per timestep, and the entire algorithm can be reformulated in terms of elementary linear algebra operations. \onlyLongVersion{For instance, the prediction step is implemented as a linear combination

\begin{equation}
\mathbf{u}_\mathrm{new} = \mathbf{u}_\mathrm{old} +M_{1} \mathbf{v}_\mathrm{old} + M_{2} \mathbf{w}_\mathrm{old} + M_{3} \mathbf{u}_\mathrm{old} + M_{4} \mathbf{u}_\mathrm{old} + \mathbf{S}\left(\mathbf{u}_\mathrm{old}\right) \nonumber
\end{equation}

where $\mathbf{u}, \mathbf{v}, \mathbf{w}$ are vectors containing temporary data computed before a prediction step of the unknowns and $M_i$ are banded matrices with four bands.
}


\section{Overview of the software structure}\label{sec:overview_software}

HONEI comprises a set of shared libraries that applications (\emph{library clients} in our terminology) can build upon. These libraries form a typical frontend-backend structure. The frontend libraries provide high-level numerical operations and container data structures such as dense, sparse and banded matrices, as well as dense and sparse vectors. Frontend libraries also provide utility functions and application-specific operations. The backend libraries provide the necessary infrastructure and low-level, hardware-specific implementations to execute the frontends computations on the different supported architectures.

For instance, the Cell backend provides mechanisms to efficiently dispatch jobs on the SPEs and perform data transfers between their local storage and main memory. These implementational details are not visible to the user. Generic template programming enables the desired functionality and lets the user select the target platform individually for each operation or globally for all operations performed in an application by applying architecture tags such as \texttt{tags::Cell} or \texttt{tags::CPU::SSE}. 

A general high-level implementation, using STL-style iterators and architecture independent parts of the C++ programming language only, provides a generic default version for all operations. This implementation does in general not receive a tremendous amount of tuning, but is guaranteed to work on all architectures that provide a reasonably modern C++ compiler.

\begin{figure}[htbp]
\centering
\includegraphics[height=0.60\textwidth,width=0.8\textwidth,keepaspectratio]{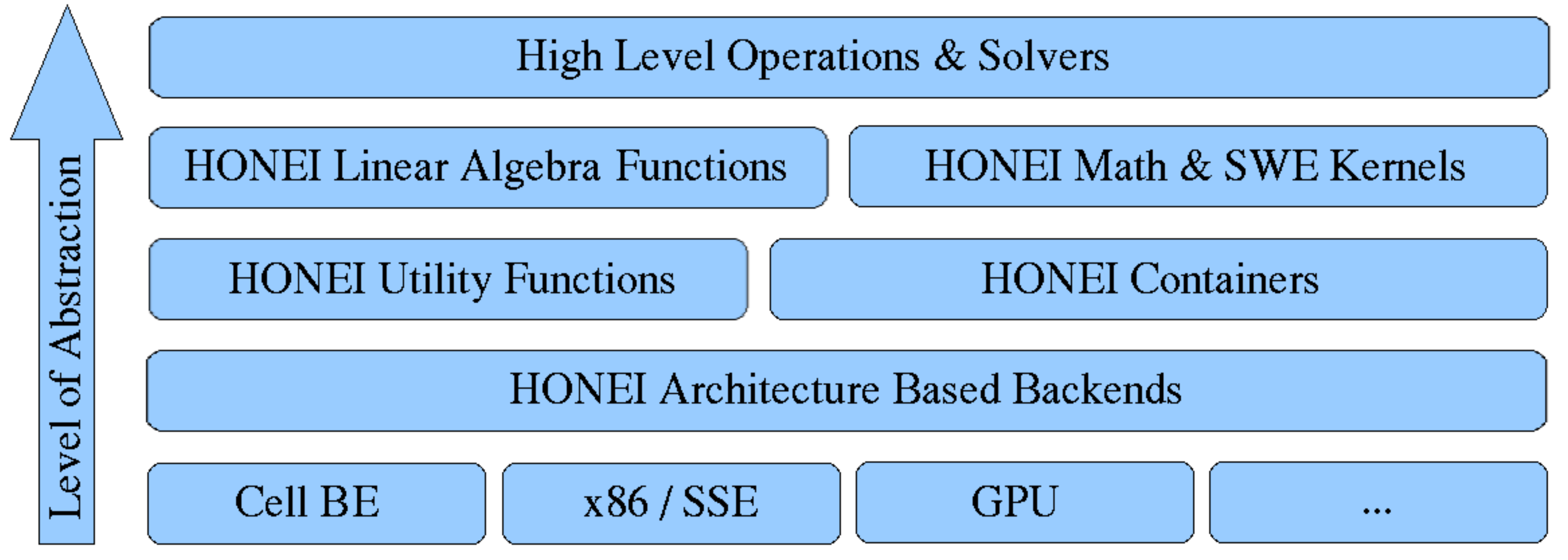}
\caption{HONEI bottom up structure and components, from the underlying hardware to the application level. }
\label{fig:HONEI_LAYOUT}
\end{figure}

Figure~\ref{fig:HONEI_LAYOUT} visualises the bottom-up structure of HONEI and the growing level of 
abstraction. Users can immediately benefit from HONEI's hardware abstraction by building applications using the templated top-level functions only. 
HONEI then automatically assures that highly tuned code is issued on the target architecture.
To faciliate fine-tuning, the configuration file \texttt{.honeirc} provides access to runtime parameters, such as the number of SPEs to use when executing Cell programs or the thread block partition for the GPU backend.
At the same time, experienced developers are free to write their own hand-tuned implementations on top of HONEI's backends addressing more application-specific optimisation
techniques.


\section{Description of the individual software components}\label{sec:description_software}

\subsection{Frontends}

The most important frontend library is \texttt{libhoneila}, HONEI's linear algebra library. It provides templated container classes for different matrix and vector types. For instance, banded matrices support both arbitrary and fixed positions of the bands, the latter ones can be exploited for improved performance of matrix-vector products. All containers follow an explicit-copy idiom, i.e., copy operations on their data will only occur by invoking a container's \texttt{copy()} method. A rich set of linear algebra operations has been implemented and specialised for the supported architectures \texttt{SSE}, \texttt{CUDA} and \texttt{Cell}, currently focussing mostly on operations needed for sparse problems.

The numerics and math library \texttt{libhoneimath} contains high performance kernels for iterative linear system solvers as well as other useful components like interpolation and approximation. While the latter are comparatively straight forward kernels, the iterative solvers are almost applications in themselves, specialised to support different preconditioning and smoothing operators as well as mixed precision schemes. Here generic programming provides the opportunity to keep interfaces clean, which in particular means that a given solver can be trivially plugged into another solver as its preconditioner. We refer to the sample code in file \texttt{honei/math/multigrid.hh} and the associated client in \texttt{clients/poisson/scenario\_controller.hh} in the HONEI tarball for details.

Application specific operations for solving the Shallow Water equations on a specific backend with the fully explicit solver described in Section~\ref{sec:swe} are included in the \texttt{libhoneiswe} library. For performance reasons, few kernels have been reimplemented and hand-tuned instead of being built via straight forward concatenation of HONEI frontend calls. These kernels are naturally highly specialised in such a way that they cannot be reused in other applications. We discuss this trade-off further in Section~\ref{sec:results_swe}.


\subsection{Backends}\label{sec:backends}

The variety of the supported architectures is reflected by the set of different backend libraries. Immediate user interaction with these backend libraries is neither intended nor guaranteed to work under all circumstances. For details on the Cell BE we refer to Pham et al., Kahle et al., Buttari et al.\ and the IBM documentation~\cite{Pham:2005:TDA,Kahle:2005:ITT,IBM_CellBE,Buttari:2007:SCO}; for a description of GPU architecture and the CUDA programming environment see Lindholm et al.\ and the CUDA documentation~\cite{Lindholm:2008:NTA,NVIDIA:2007:CUDA}.

The current implementation does not yet virtualise memory. This means that the limited amount of device memory poses a hard restriction that the application programmer has to keep in mind. However, HONEI exits cleanly if more memory than available is requested.

\texttt{libhoneibackendssse}, selected by the tag \texttt{tags::CPU::SSE}, provides fast SSE2 implementations of the most important frontend functions like norms or matrix-vector products. We rely on the SSE built-ins of the Intel C++ Compiler, which are also supported by the GNU C++ Compiler. 

\texttt{libhoneibackendscell}, selected by the tag \texttt{tags::Cell}, is a compound library comprising implementations of the most important frontend functions as well as a supporting build-system to create SPE programs (\emph{kernels}) at runtime. These kernels are written in C++, however, in order to not bloat the limited SPE memory which needs to hold both instructions and data, we decided to disable support for RTTI and exception handling, resulting in a memory footprint comparable to plain C programs while still supporting C++ features like templates.
This library additionally provides a remote procedure call (RPC) system to relay function calls from the PPE to the SPEs. It relies on \texttt{libspe2}, a library provided by IBM~\cite{libspe} and some tools provided by the IBM Cell BE SDK 2.0~\cite{Perrone:2006:CBS}.
Last but not least, there are templates to facilitate writing of new callable functions and registering them with the RPC system.
When a given operation is passed to the Cell backend, a new instruction (consisting of an opcode and the operation parameters) is generated and stored in an instruction queue.
If one of the available and idle SPEs has previously been configured to process this opcode (and hence is running a kernel that is able to execute the instruction), the scheduler wakes it up.
Otherwise it signals an idle SPE to load the necessary kernel.
The activated SPE pulls the instruction and the corresponding data from main memory, processes the instruction and finally stores the result to main memory again.
With this approach, we achieve good load distribution among the SPEs and disburden the PPE's DMA controller by distributing all DMA transfers to the SPE's DMA controllers. The Cell backend has been successfully tested on the PlayStation 3 and on QS20 and QS22 blade systems by IBM.

\texttt{libhoneibackendscuda}, selected by the tag \texttt{tags::GPU::CUDA}, is based on NVIDIA's CUDA programming environment~\cite{NVIDIA:2007:CUDA} in version 2.0. The GPU device portion of the library comprises the same set of functions as those collected in \texttt{libhoneibackendssse}. The device code is written in an extension of plain C and compiled with the CUDA compiler \texttt{nvcc}. To avoid unnecessary memory transfers from main memory to the GPU's device memory or vice versa, we implemented a transparent memory scheduler, which decides if processed data needs to be transferred. For details on CUDA programming, we refer to the programming guide~\cite{NVIDIA:2007:CUDA} and previous work~\cite{Goeddeke:2008:PAA}.

\subsection{Client applications}\label{sec:clients}

The \texttt{clients} directory finally stores client applications built on top of the HONEI libraries. In the scope of this paper, the clients \texttt{honei-poisson} and \texttt{honei-swe} solve the Poisson problem (Section~\ref{sec:poisson}) and perform SWE simulations (Section~\ref{sec:swe}) respectively.


\section{Installation instructions}
\label{installation}
HONEI is installed through the usual GNU toolchain. For details, we refer to the README file provided in the installation tarball.


\section{Results}\label{sec:results}

In this section, we present results for the two example applications introduced in Section~\ref{sec:background}. Besides performance, we are particularly interested in evaluating HONEI's library approach; one of the most important aspects here is to assess the benefits and limitations of the concept of hardware abstraction on the level of individual applications. 

For the CPU computations we use two common x86 machines, an Intel Core2Duo 6320 (1.86~GHz, 4~MB shared L2 cache) and an AMD Opteron X2 2214 (2.2~GHz, 1~MB L2 cache per core). Code is compiled with gcc version 4.1.2 and compiler settings tuned for each machine. In this paper we only use a single core per CPU; a multi-core CPU backend is currently being developed. All Cell results are obtained on the Sony PlayStation 3 (PS3) running Yellow Dog Linux 4, and we use an NVIDIA GeForce 8800 GTX graphics board included in the Opteron machine for the GPU tests. It is worth noting that the \emph{usable} device memory on the PS3 amounts to only 200\,MB due to the needs of the operating system, which significantly inhibits the tests we can perform. In contrast, more than 700\,MB are available on the GPU.


\subsection{Basic architecture comparison}\label{sec:blas}
\begin{figure}[hb]
\begin{center}
\includegraphics[height=5cm]{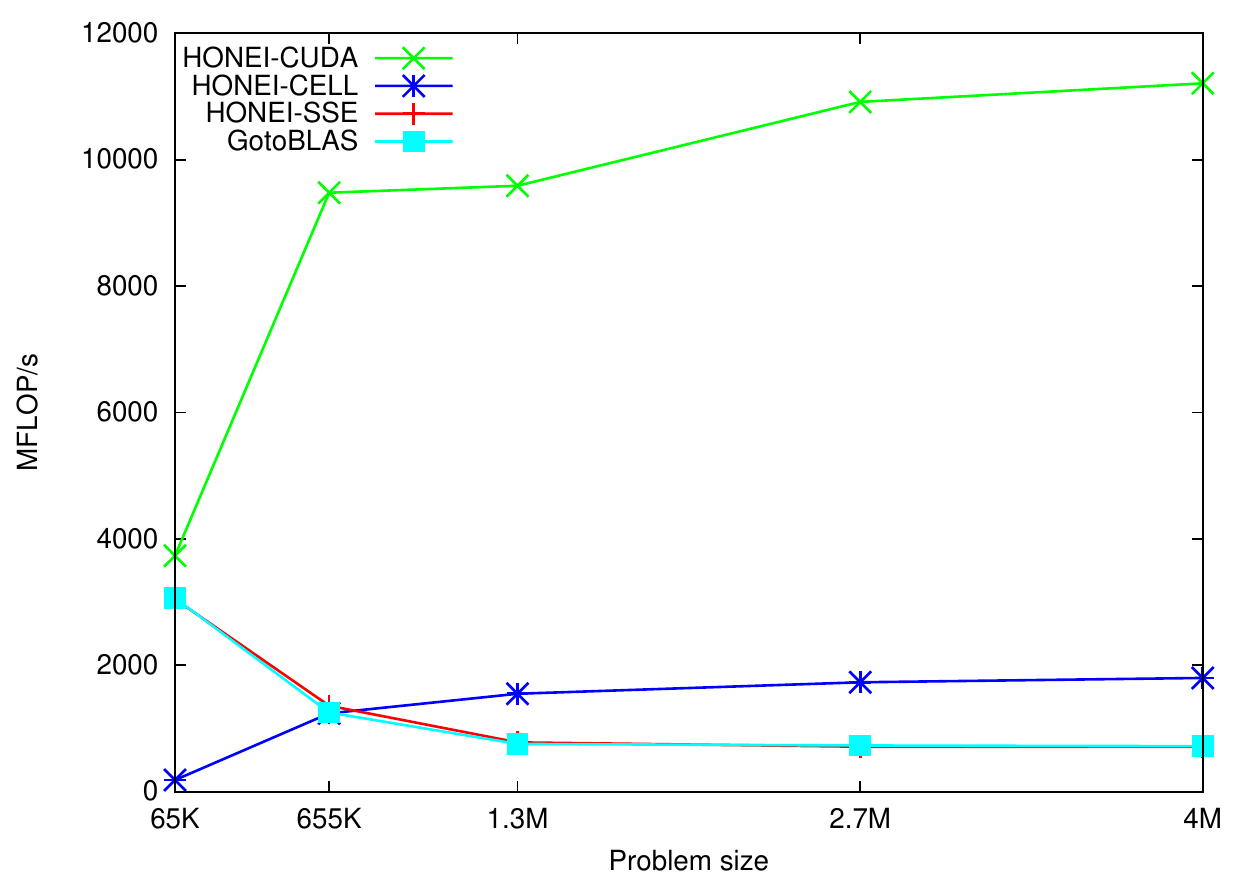}
\end{center}
\vspace*{-0.5cm}\caption{Performance of an \texttt{axpy} kernel with different HONEI architectures and GotoBLAS.}
\label{mflops_axpy}
\end{figure}

We first demonstrate low level benchmarks (see Figure \ref{mflops_axpy}) for the standard \texttt{saxpy} operation, and observe a 16-fold (GPU) and 2.5-fold (PS3 using 4 SPEs) speedup versus the Core2Duo system, which reflects the available off-chip bandwidth, the decisive performance factor for this purely memory bound operation. On CPUs, performance drops once the data does not fit into cache anymore, while on the Cell and on GPUs, performance only surges once the problem sizes are large enough to hide the configuration overhead associated with these devices.

These experiments also confirm the asymptotical \lq optimality\rq\ of HONEI's SSE backend for vector-vector operations, as we achieve exactly the same performance level as (single threaded) GotoBLAS~\cite{GotoBLAS}, an SSE assembly BLAS implementation that has been extremely tuned with respect to cache and TLB blocking. We note that we achieve similar speedups for other BLAS level 1 operations.


\subsection{Multigrid Poisson solver}

To solve the Poisson equation, a fundamental model problem in many application domains (see Section~\ref{sec:poisson}), we employ HONEI's generic multigrid solver, implemented in \texttt{libhoneimath}. We configure the solver to perform a $V$ cycle with two damped pre- and post\-smooth\-ing Jacobi steps, and use an unpreconditioned Conjugate Gradient solver for the coarse grid problems. \onlyLongVersion{The cheapest multigrid cycle suffices for this configuration.} Internally, the solvers call basic low level linear algebra operations from \texttt{libhoneila}, and therefore this experiment is well-suited to assess how low-level performance translates to the application level, and thus HONEI's generic library approach in general. The application itself is implemented once, and hardware specialisation is achieved transparently to the user by setting the corresponding architecture and precision tags (see Section~\ref{sec:overview_software} for details). 

As the supported architectures compute much faster in single than in double precision (SSE, Cell) or do not support double precision at all (8800~GTX GPU), we employ a (slightly more costly) mixed precision solution scheme~\cite{Goeddeke:2007:PAA}. The basic idea is to execute a simple defect correction scheme in double precision on the CPU, which is preconditioned by performing two multigrid cycles (which amounts to gaining one digit in the residual) in single precision on a co-processor. In this test, we use HONEI's SSE and CUDA backends on the Opteron machine.

\subsubsection{Accuracy}
Assessing the accuracy of our mixed precision scheme is very important. We use the analytically known Laplacian of a polynomial test function (with pure Dirichlet boundary conditions) as right hand side to the Poisson equation. We thus know the exact analytical solution, and can compare the computed results in the integral $L_2$ norm. For the bilinear nonconforming Finite Elements that we use for the discretisation, the $L_2$ error reduces by a factor of four per mesh refinement step ($h^2$) provided that the computation is performed with sufficient floating point precision. Table~\ref{tab:errorreduction} summarises our results; recall that refinement level $L$ yields a problem size of $(2^L+1)^2$ degrees of freedom. 

\begin{table}[hbt]
\footnotesize
\centering
\begin{tabular}{|r|l|c|l|c|l|c|}
\hline
 & \multicolumn{2}{c|}{single precision} & \multicolumn{2}{c|}{double precision} & \multicolumn{2}{c|}{mixed precision} \\
level&$L_2$ error & red.& $L_2$ error & red. & $L_2$ error& red.\\
\hline
2  &5.1845E-5 & -   &5.1843E-5 & -   &5.1843E-5 & - \\
3  &1.2121E-5 &4.28 &1.2124E-5 &4.27 &1.2124E-5&4.28 \\
4  &2.9944E-6 &4.05 &2.9748E-6 &4.07 &2.9748E-6&4.08 \\
5  &1.0683E-6 &2.80 &7.4013E-7 &4.02 &7.4013E-7&4.02 \\
6  &1.5614E-6 &0.68 &1.8481E-7 &4.00 &1.8481E-7&4.00 \\
7  &5.7570E-6 &0.27 &4.6188E-8 &4.00 &4.6188E-8&4.00 \\
8  &2.7530E-2 &0.00 &1.1546E-8 &4.00 &1.1546E-8&4.00 \\
9  &3.8110E-1 &0.07 &2.8864E-9 &4.00 &2.8761E-9&4.01 \\
10 &6.2065E0 &0.06 &7.2134E-10 &4.00 &7.1745E-10&4.01 \\
\hline
\end{tabular}
\caption{Accuracy tests of the multigrid Poisson solver: Reduction of $L_2$ errors in various precision formats.}
\label{tab:errorreduction}
\end{table}

We observe that computing entirely in double precision is accurate enough for this test problem, the $L_2$ errors reduce as expected. In single precision\footnote{The $L_2$ error is of course computed in double precision.} however, the results are completely wrong, already for the small problem size of 1089 unknowns (level 5): Increasing the level of refinement and hence the problem size even increases the error again. The rightmost columns in Table~\ref{tab:errorreduction} show that the mixed precision scheme achieves exactly the same accuracy as computing entirely in double precision.

\subsubsection{Performance}

\begin{figure}[ht]
\begin{center}
\includegraphics[height=5cm]{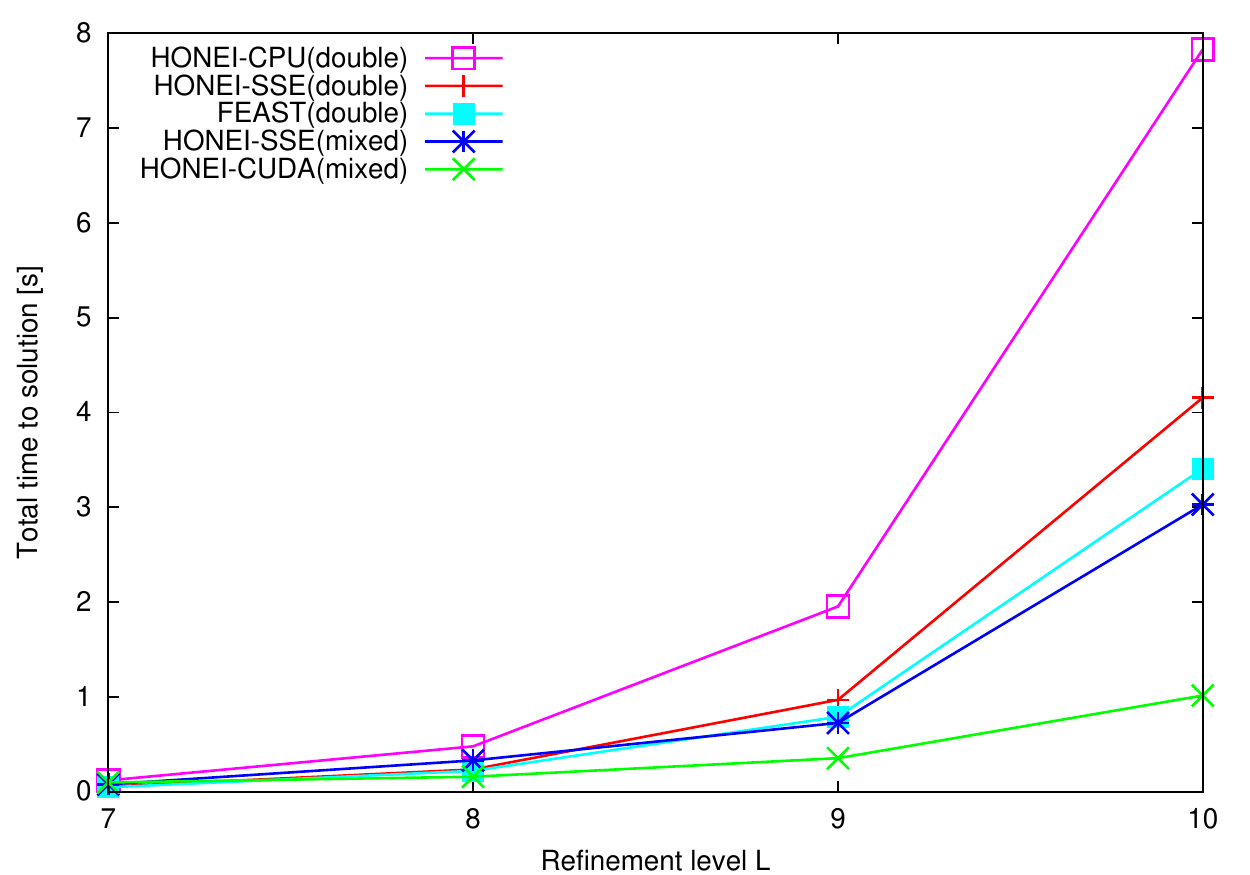}
\end{center}
\vspace*{-0.5cm}\caption{Performance tests of the multigrid Poisson solver on the Opteron system and the GPU.}\label{fig:performance_poisson}
\end{figure}

To demonstrate the flexibility of HONEI's multigrid solver, we evaluate performance based on a Poisson problem with mixed (free) Neumann and (geometric) Dirichlet boundary conditions. To assess the absolute performance, we perform the same experiment with a second Finite Element toolkit, FEAST~\cite{Becker:2007:SUM}, which is extremely tuned for this problem. Figure~\ref{fig:performance_poisson} summarises our performance evaluation.

We first observe that in double precision, the HONEI SSE solver is slightly slower than FEAST. For the larger problem sizes, FEAST is typically 20\% faster due to its more aggressive exploitation of the banded matrix structure. Comparing HONEI's default CPU implementation that relies on compiler optimisations alone with the double precision SSE implementation, we observe a speedup by a factor of two on the highest refinement levels. HONEI's mixed precision SSE implementation executes more than $1/3$ faster than the double precision SSE version and outperforms FEAST, while delivering the same result accuracy. This last observation is important, as mixed precision is essentially for free in HONEI due to its templated software design, in contrast to the established, Fortran-based FEAST toolkit.

For the trivial problem sizes not included in Figure~\ref{fig:performance_poisson}, the GPU accelerated solver is slower than the SSE version which computes entirely in cache as observed in the previous experiment. For the largest problem instances, we achieve speedup factors of $3$ versus the mixed precision SSE implementation and $4$ versus the double precision SSE version, in line with expected performance. The limiting factor in these experiments is the expensive data transfer over the PCIe bus within the mixed precision scheme. Newer GPUs support native double precision, removing this bottleneck. These timing measurements confirm that for a solver constructed entirely from low-level operations, HONEI's library approach works extremely well and hardware abstraction on the level of individual operations translates directly to application speedup. In addition, the hardware oriented approach provides high performance kernels completely independent of the compiler in use.


\subsection{Shallow Water solver}\label{sec:results_swe}

\begin{figure}
\begin{center}
\includegraphics[height=0.2\textheight]{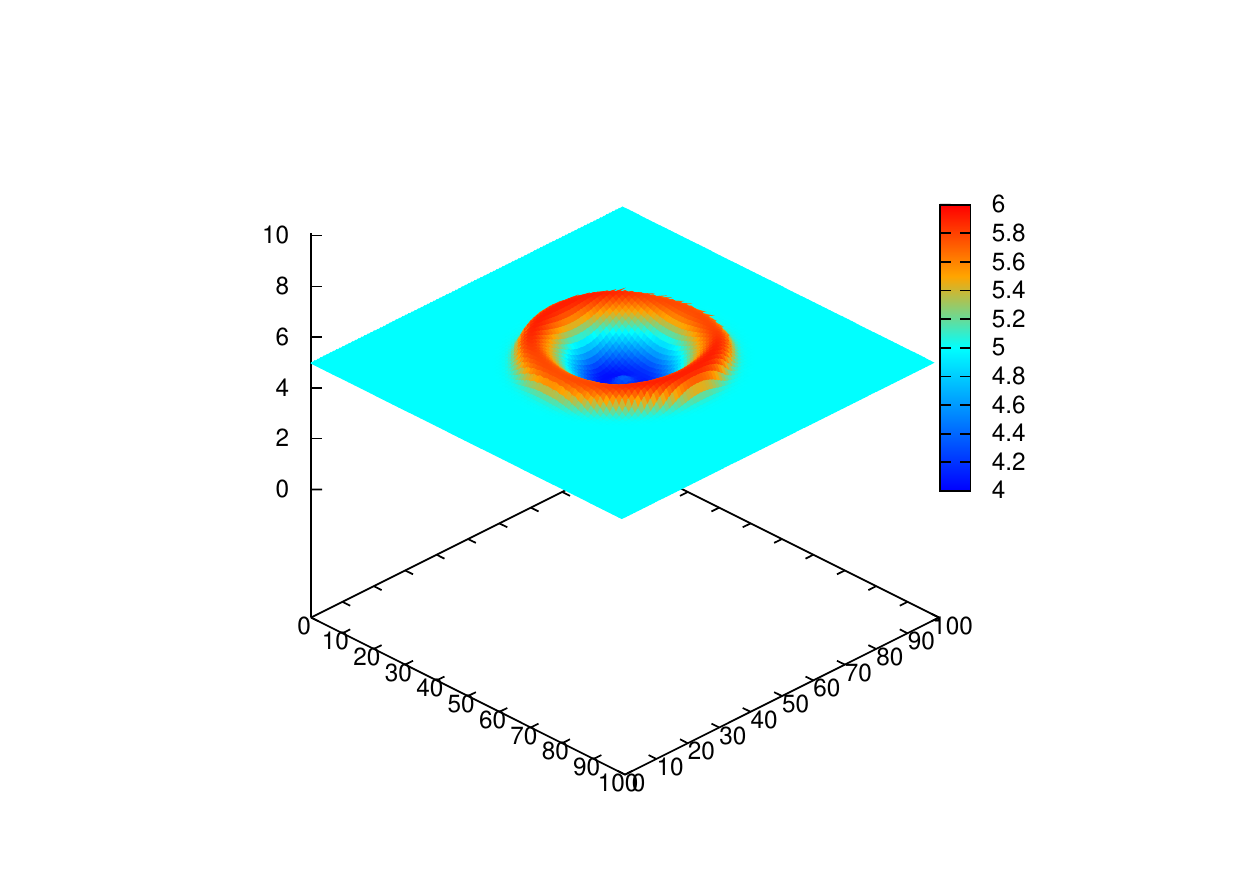}\includegraphics[height=0.2\textheight]{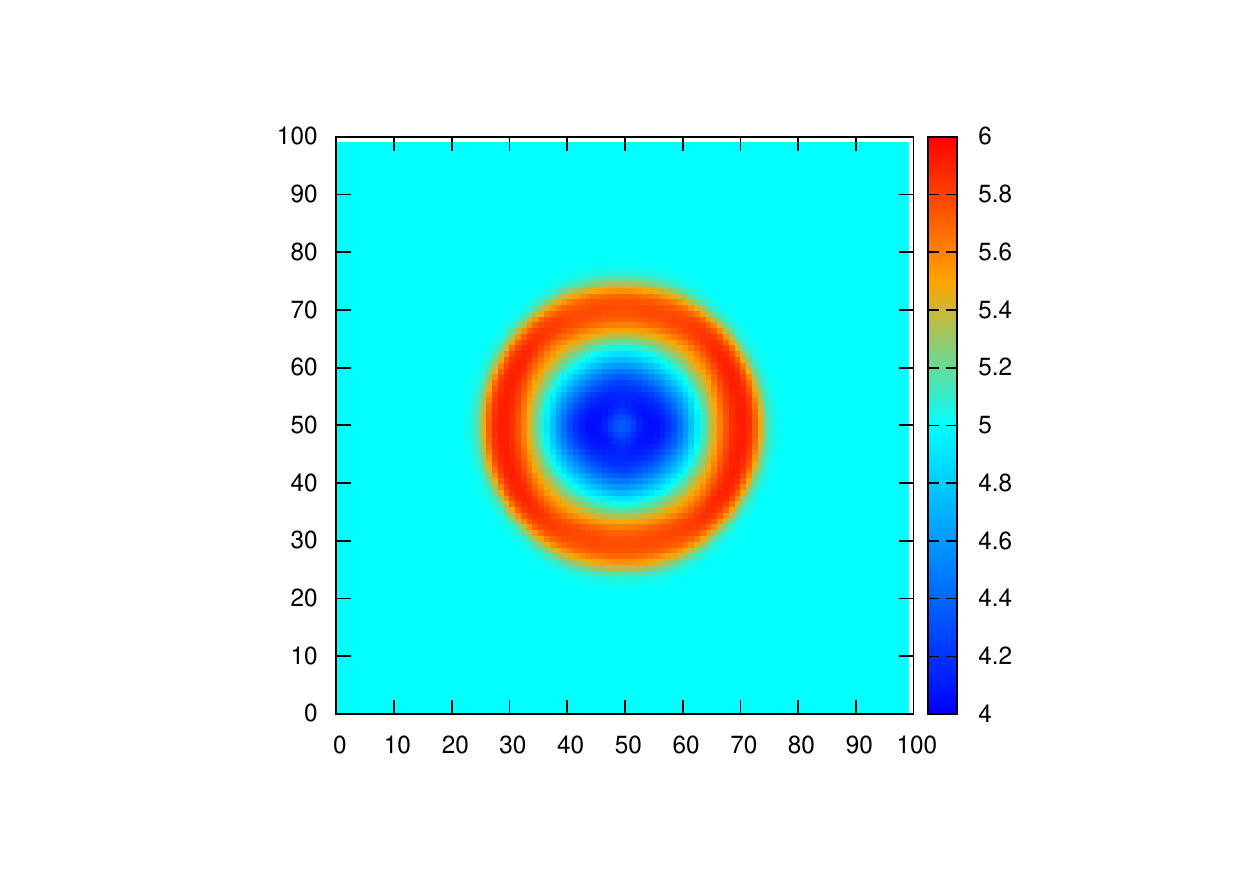}\\
\vspace*{-2em}
\includegraphics[height=0.2\textheight]{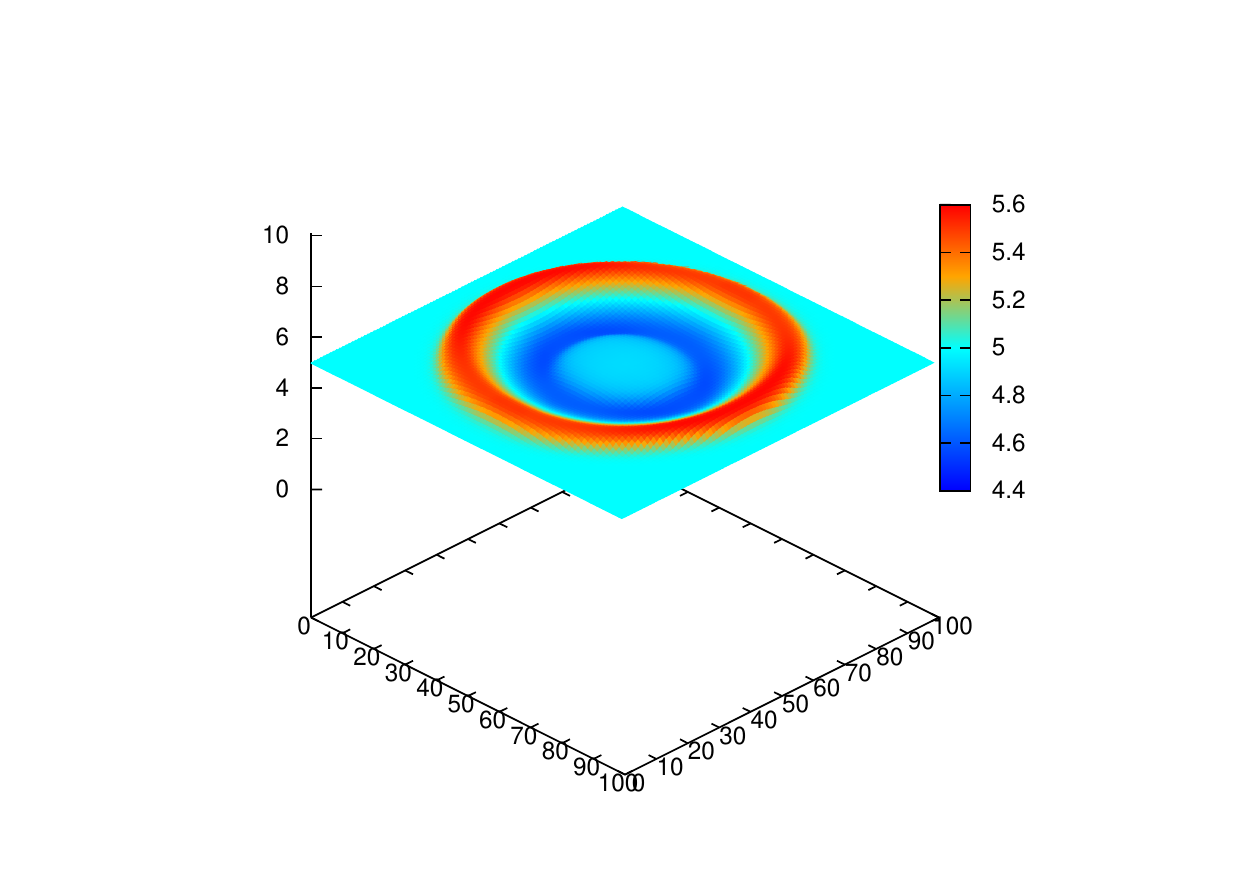}\includegraphics[height=0.2\textheight]{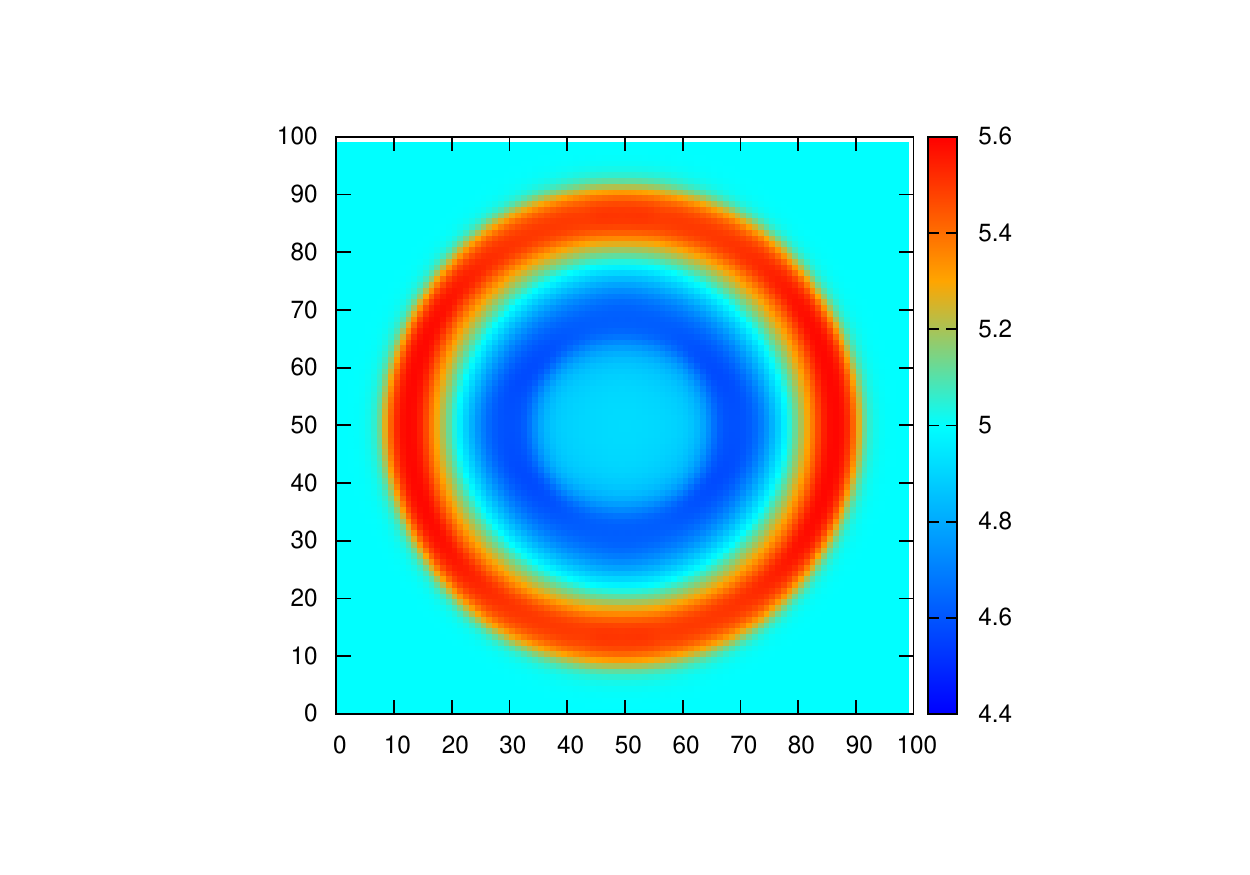}\\
\vspace*{-2em}
\includegraphics[height=0.2\textheight]{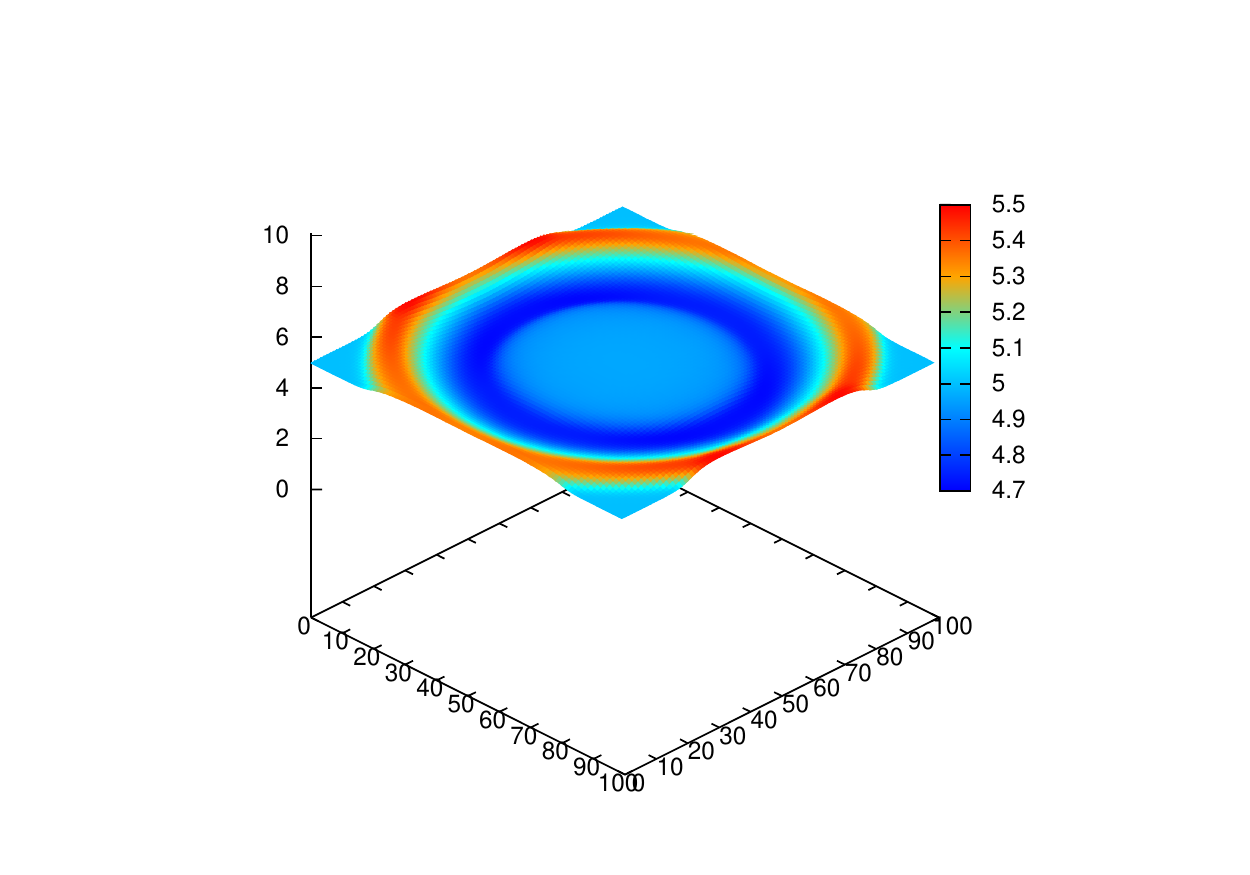}\includegraphics[height=0.2\textheight]{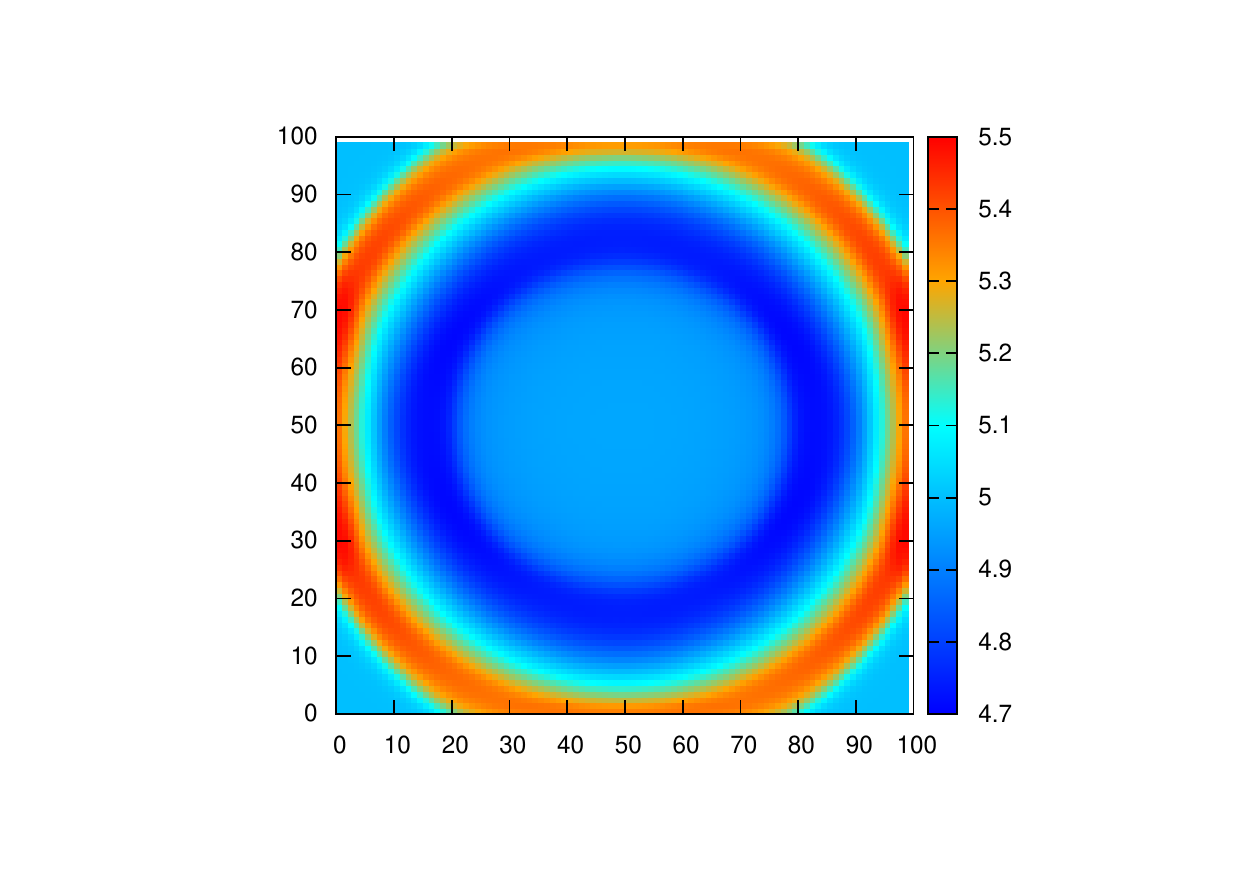}\\
\vspace*{-2em}
\includegraphics[height=0.2\textheight]{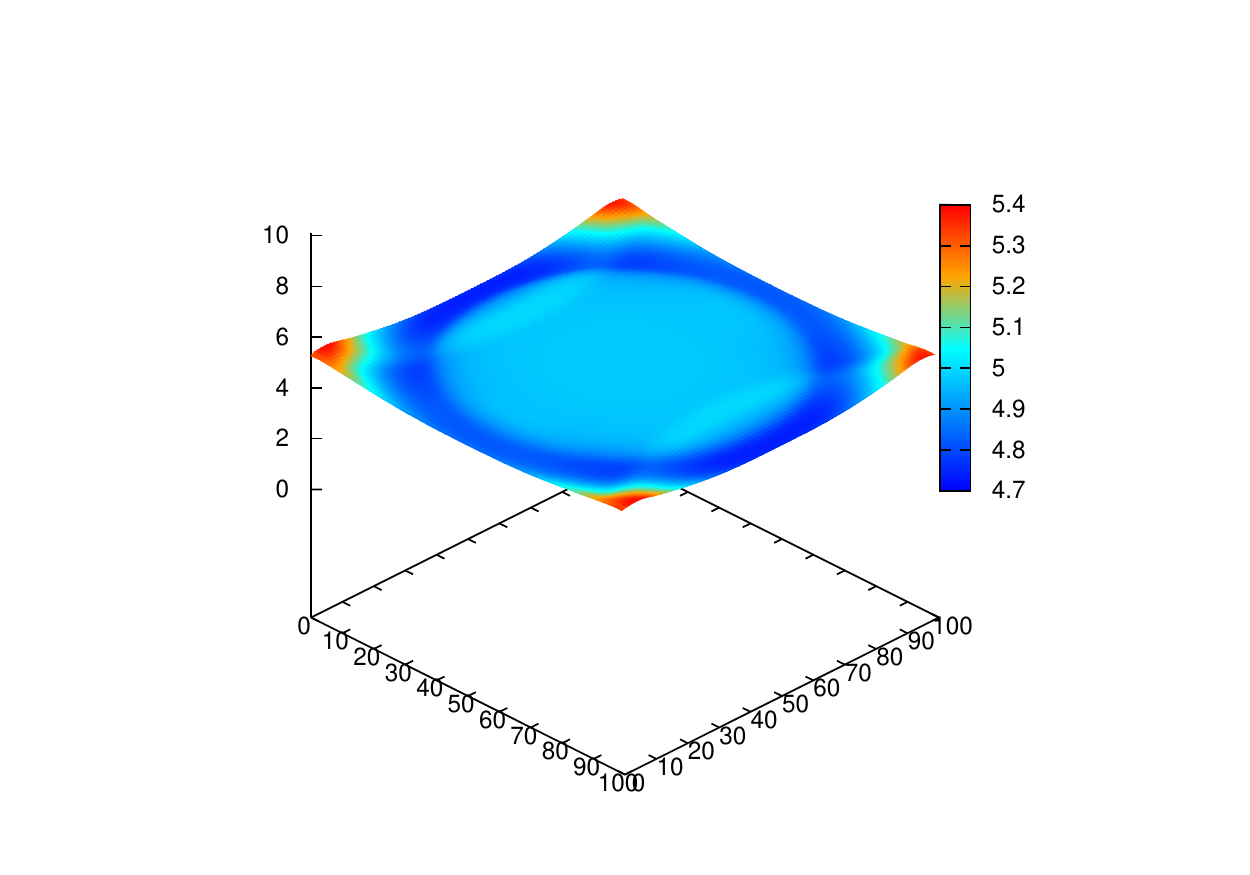}\includegraphics[height=0.2\textheight]{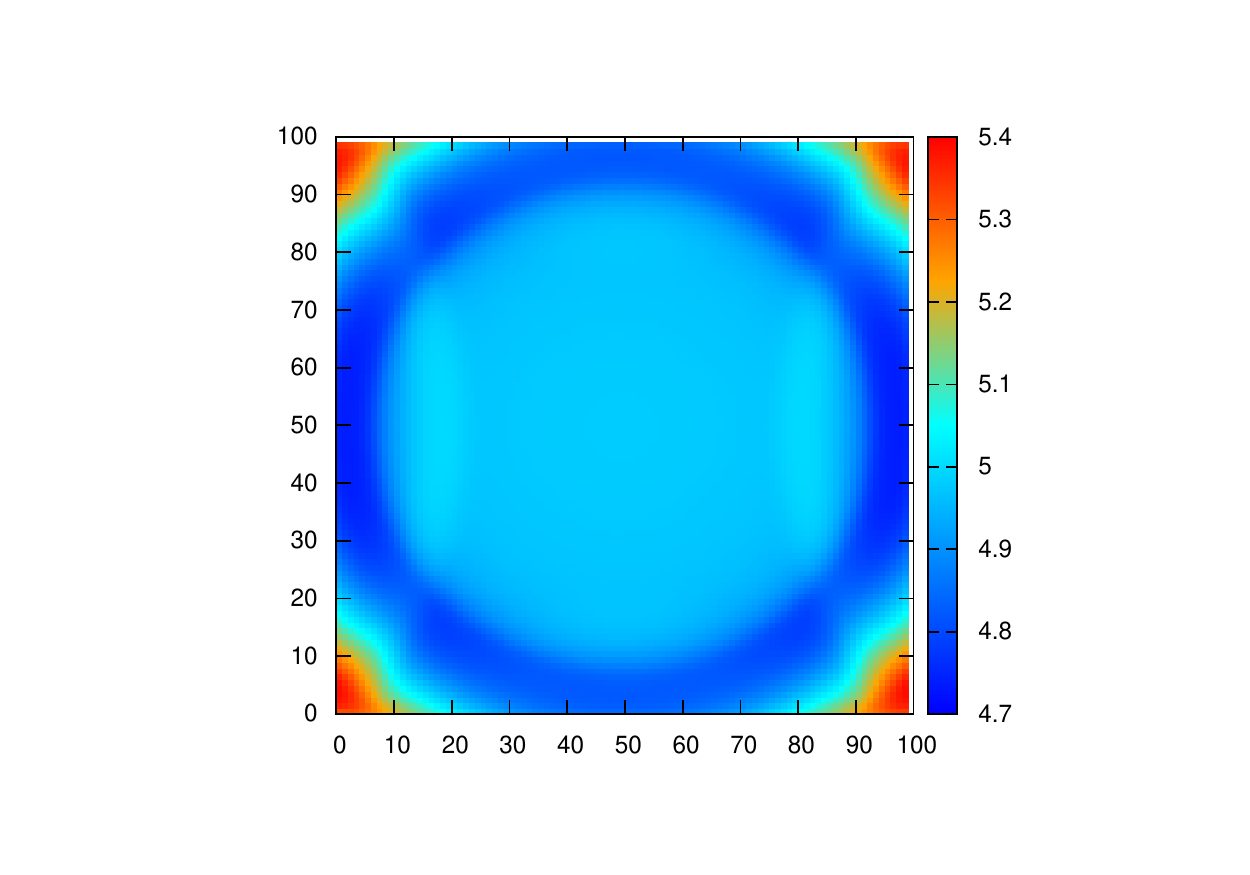}\\
\end{center}
\caption{Circular dambreak simulation on a $100\times100$ grid computed with our SWE solver on the PS3. Top down: Water height after $t=10.42s$, $t=20.83s$, $t=31.25s$ and $t=46.6s$. Left column shows 3D view and right column the associated height map. $\Delta x = \Delta y = 5m$.}\label{fig:swe-picture}
\end{figure}

We execute our SWE solver on the Core2Duo machine and the PlayStation 3. Figure~\ref{fig:swe-picture} shows a typical circular dambreak simulation. 

\subsubsection{Application-specific optimisations}

\begin{figure}[htb]
\begin{center}
\includegraphics[height=5cm]{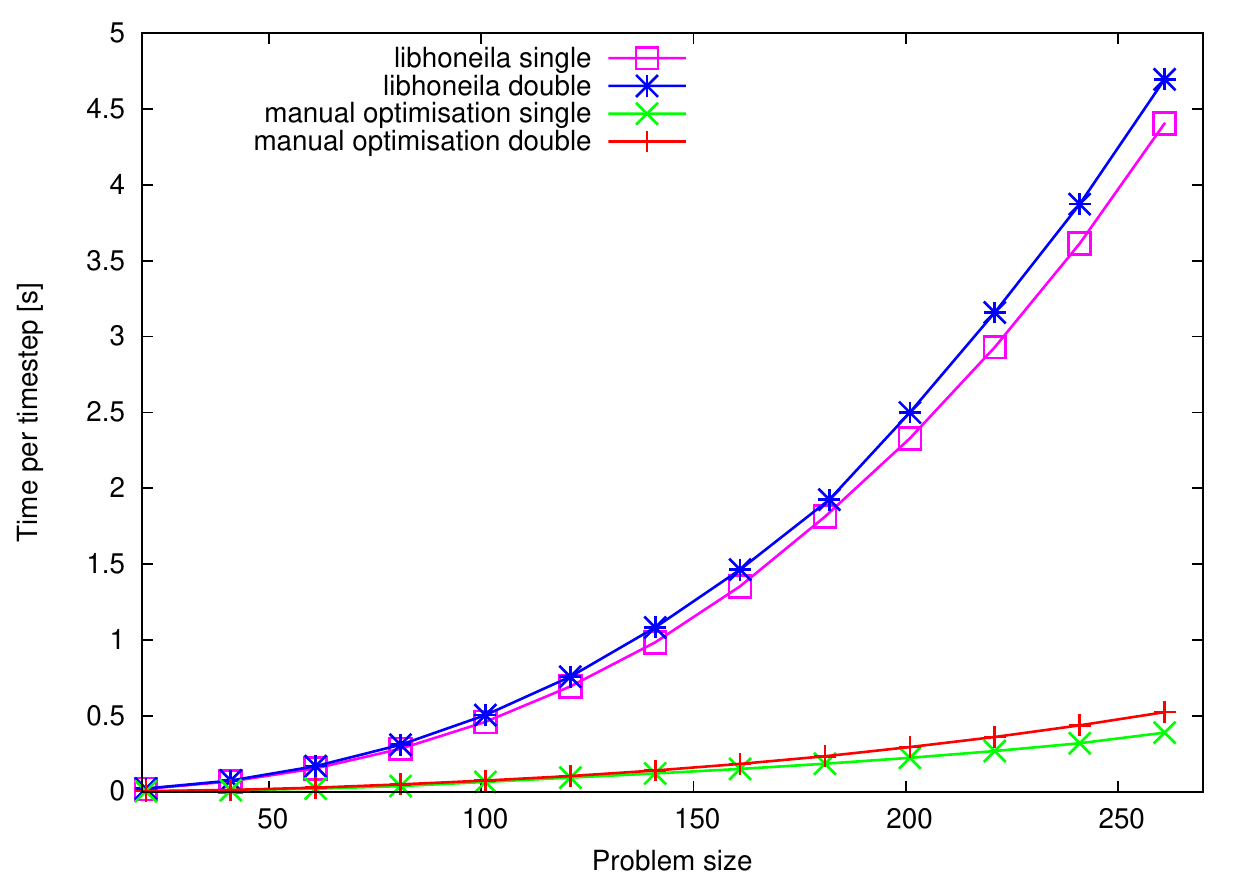}
\end{center}
\vspace*{-0.5cm}\caption{SWE solver performance comparison: Hand-crafted SSE application kernels vs. compiler-crafted ones on the Core2Duo.}\label{fig:versus}
\end{figure}

Our initial implementation used a concatenation of HONEI linear algebra operators for matrix and vector assembly (which is done once per timestep), using the default CPU backend without SSE optimisations. Detailed analysis revealed that almost 50\% of the execution time is spent in these stages of the algorithm, and the compiler is not able to optimise the code properly due to irregular memory access patterns. We thus implemented manually optimised SSE and Cell versions of these kernels as an example of application-specific tuning. Figure~\ref{fig:versus} shows the effect of this optimisation. Note that the difference between the default CPU implementation and the SSE version increases significantly with the problem size and is much higher than for the Poisson solver due to the abovementioned irregular memory access patterns. The results clearly highlight the impact of such efforts, especially if the percentage of application specific code is quite high. Application programmers are therefore encouraged to take advantage of the infrastructure provided by the HONEI backends to implement hardware-specific application kernels if necessary. 

The HONEI tutorial illustrates in detail how such application-specific kernels are created, using the norm of a generalised residual calculation based on the same banded matrix container as needed by the Poisson solver: $s = ||\alpha \mathbf{y} + \beta A \mathbf{x}||_2$. This composite kernel is not part of HONEI's linear algebra library, only the three atomic operations it comprises are. In particular, the code examples in the tutorial allow to assess the implementational effort to realise application-specific kernels using the infrastructure HONEI provides.

\subsubsection{Accuracy and performance on the Cell processor}

As peak single precision arithmetics on the 
Cell processor is roughly 14 times faster than double precision performance, we have designed two novel mixed 
precision configurations for the solver to exploit this performance difference: Our first idea was to execute every $k$-th iteration in double 
precision, and employ single precision otherwise. We then analysed the 
internal structure of one timestep in more detail and identified the stages where double precision has the highest 
effect on the accuracy of the results. We found that the predictor/corrector scheme is most crucial for final 
result accuracy, and our resulting final mixed precision scheme executes the prediction step in double precision 
and everything else in single precision. 

\begin{figure}[htb]
\begin{center}
\includegraphics[height=5cm]{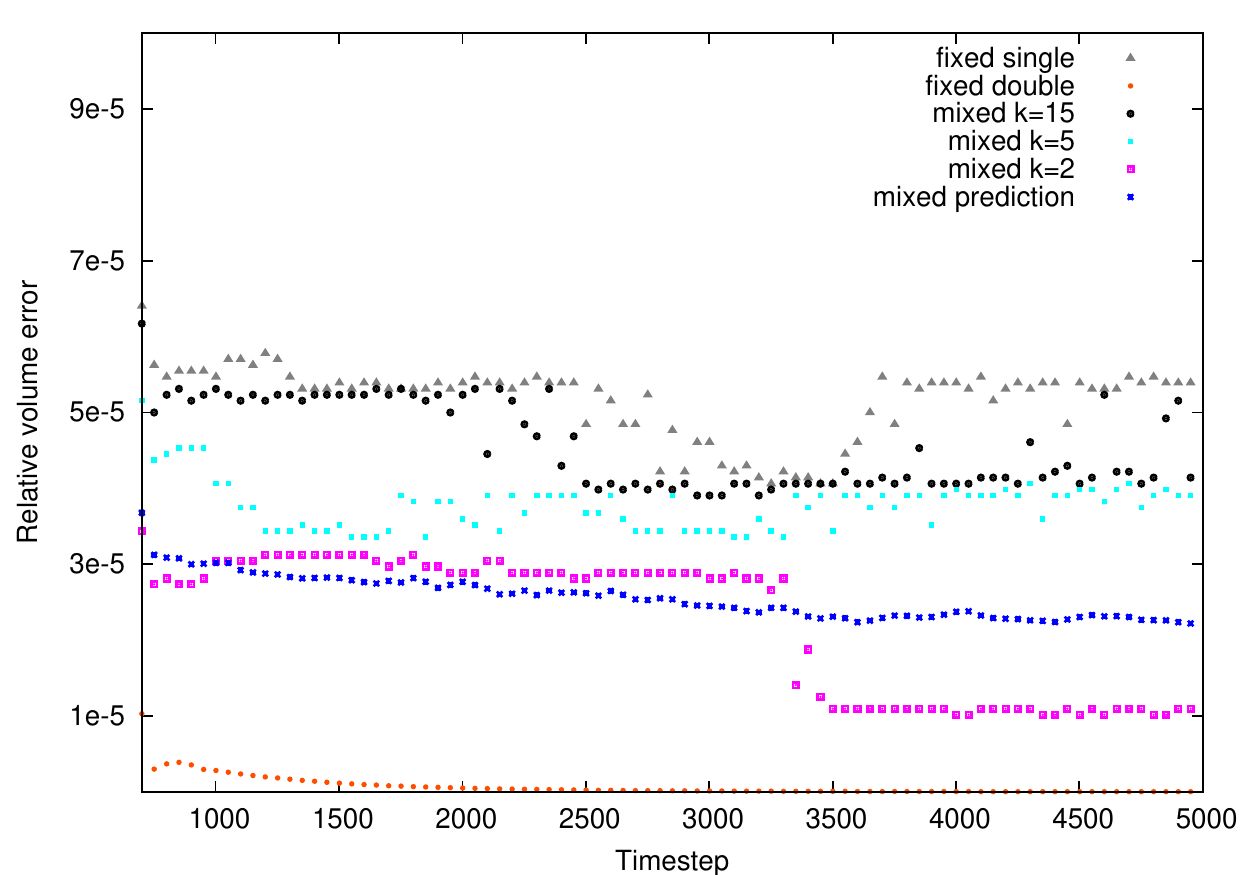}
\end{center}
\vspace*{-0.5cm}\caption{Relative volume error for different fixed and mixed precision SWE solver configurations on the PS3 on a $100\times 100$ grid.}\label{fig:mass}
\end{figure}

\begin{figure}[htb]
\begin{center}
\includegraphics[height=5cm]{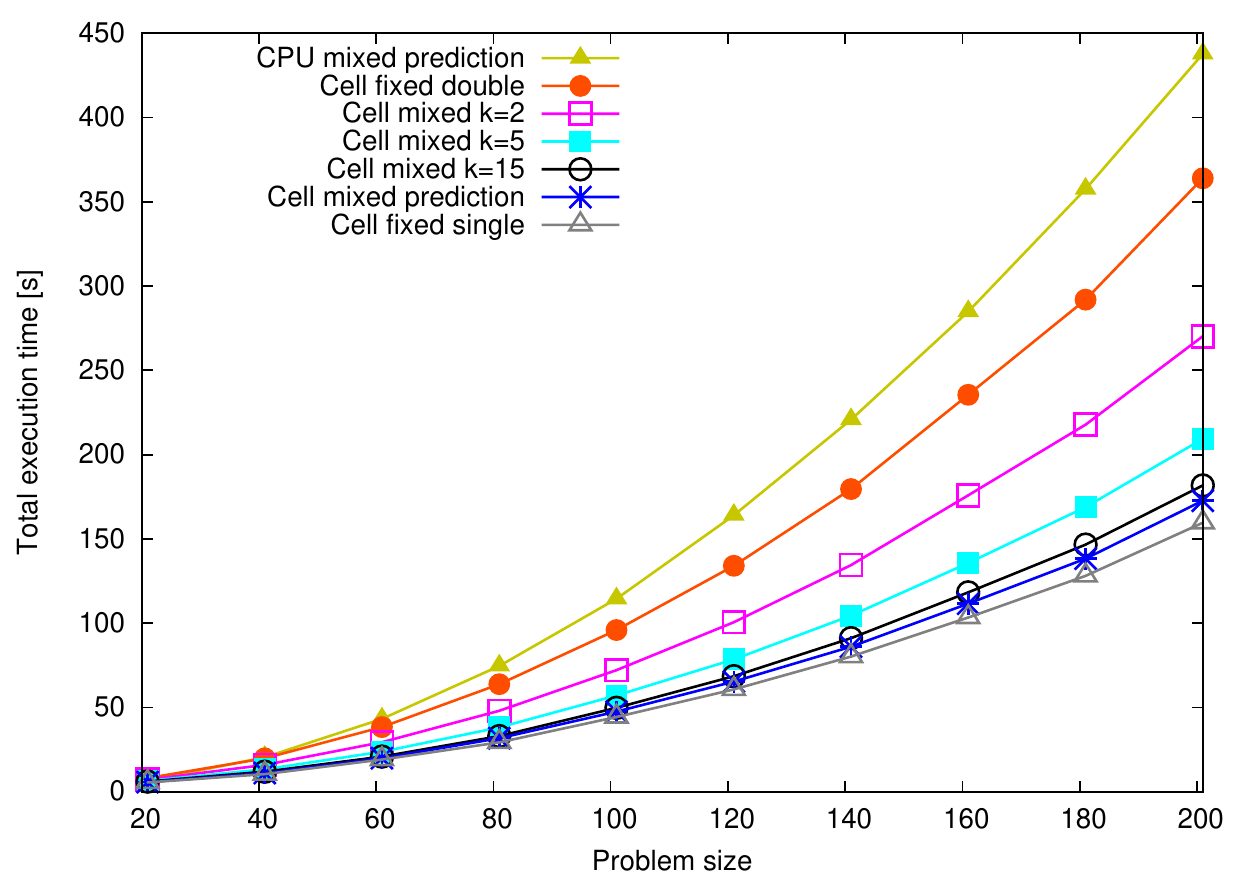}
\end{center}
\vspace*{-0.5cm}\caption{Total execution time of the SWE solver for different fixed and mixed precision configurations on the PS3 and the Core2Duo system.}\label{fig:mixedbench}
\end{figure}

As no analytic solution is available, we examine the solver's mass conservation capabilities to assess the impact of varying precision on the achievable accuracy. Figure~\ref{fig:mass} shows the relative volume error measured at fixed timesteps during the simulation. The fixed single and double precision results differ significantly, and the mixed precision configurations (double precision timesteps with $k \in \{2,5,15\}$ and the double precision prediction method) are distributed between them, as expected. The best result is achieved by employing double precision for the prediction step, and even though format conversion and transfer have to be performed twice per iteration (each timestep consists of two prediction and correction steps), this method is also the fastest mixed precision configuration (see Figure~\ref{fig:mixedbench}). To be more precise, while the latter method is not significantly slower than solving with fixed single precision, it halves the error. Executing the entire solver in double precision is more accurate, but much slower. Figure~\ref{fig:mixedbench} also illustrates the performance gains of our Cell implementation vs.\ the fastest CPU version using HONEI's SSE backend, we obtain a speedup factor of 2.5. These results are obtained with the fast, manually optimised assembly routines.

\section{Conclusions and future work}

We have presented HONEI, a collection of libraries that simplify programming for both conventional CPU designs and nonstandard emerging architectures. In particular, the user can program an application once, and then compile it for execution on the SIMD SSE units of CPUs, CUDA-enabled GPUs or Cell processors. 

For applications and components like linear system solvers, this hardware abstraction works particularly well, and microbenchmark performance translates directly to acceleration on the application level. However, not all applications can benefit immediately from tuned basic operations, we have demonstrated this with a solver for the Shallow Water equations. The fusion of general basic routines with application specific kernels (which lack generality, but which can often be reused in the same global context like CFD) therefore remains a future-proof concept especially when the increasing discrepancy between hardware development compared to compiler development is taken into account. In this case, the infrastructure provided by HONEI's backend libraries simplifies hardware-aware programming to a great extend.

HONEI is being continuously developed. Both the CUDA backend and the Cell backend will be extended in the near future by new and further optimised kernels. In addition, a Lattice Boltzmann Method (LBM) based application for solving the SWE and the Navier Stokes equations focusing on very fast and robust real-time computations is being implemented. Finally, with respect to emerging manycore architectures like Intel's Larrabee, a pthread-based multicore backend and a MPI backend are in early beta stage.

HONEI is available as Open Source under the GPL licence and developer snapshots of the source code can be obtained via \url{http://www.honei.org}.


\section*{Acknowledgements}
Parts of this work were supported by the German Science Foundation (DFG), project TU102/22-1. We thank all participants of PG512 at TU Dortmund for initial support. Thanks to NVIDIA for donating hardware.


\small
\bibliography{references}

\end{document}